\documentclass[11pt,twoside]{article}
\usepackage{graphicx}
\usepackage{subfigure}
\usepackage{fancyheadings}
\usepackage{amsmath}
\usepackage{amssymb}
\usepackage{cite}

\begin{document}
\newcommand{\pst}{\hspace*{1.5em}}

\newcommand{\rigmark}{\em Journal of Russian Laser Research}
\newcommand{\lemark}{\em Volume 35, Number 3, 2014}

\newcommand{\be}{\begin{equation}}
\newcommand{\ee}{\end{equation}}
\newcommand{\bm}{\boldmath}
\newcommand{\ds}{\displaystyle}
\newcommand{\bea}{\begin{eqnarray}}
\newcommand{\eea}{\end{eqnarray}}
\newcommand{\ba}{\begin{array}}
\newcommand{\ea}{\end{array}}
\newcommand{\arcsinh}{\mathop{\rm arcsinh}\nolimits}
\newcommand{\arctanh}{\mathop{\rm arctanh}\nolimits}
\newcommand{\bc}{\begin{center}}
\newcommand{\ec}{\end{center}}

\thispagestyle{plain}

\label{sh}

\begin{center} {\Large \bf
\begin{tabular}{c}
NEW INEQUALITIES FOR QUANTUM VON NEUMANN

\\[-1mm]
AND TOMOGRAPHIC MUTUAL INFORMATION
\end{tabular}
 } \end{center}

\bigskip

\bigskip

\begin{center} {\bf
V.I. Man'ko$^{1}$ and L.A. Markovich$^{2*}$
}\end{center}

\medskip

\begin{center}
{\it
$^1$P.N. Lebedev Physical Institute, Russian Academy of Sciences\\
Leninskii Prospect 53, Moscow 119991, Russia

\smallskip

$^2$Institute of Control Sciences, Russian Academy of Sciences\\
Profsoyuznaya 65, Moscow 117997, Russia
}
\smallskip

$^*$Corresponding author e-mail:~~~kimo1~@~mail.ru\\
\end{center}

\begin{abstract}\noindent
Entropic inequalities related to the quantum mutual information for bipartite system and tomographic mutual information
is studied for Werner state of two qubits. Quantum correlations corresponding to entanglement properties of the qubits in Werner state
are discussed.
\end{abstract}

\medskip

\noindent{\bf Keywords:} Entropic inequalities, quantum information, Werner state, tomographic probabilities, qudit.

\section{Introduction}
The two-qubit systems can demonstrate quantum correlations and these correlations correspond to entanglement
phenomenon \cite{schredinger:35} or to the violation of Bell inequalities \cite{Horn}. Also the correlations can be
associated with quantum discord \cite{Yurkevich,Mscord}. 
The quantum discord is related to difference of classical Shannon  information  behavior \cite{Shannon} and quantum
information behavior determined by von Neumann entropy of a composite bipartite systems and the entropies of its subsystems.
Recently \cite{Dodonov,OVManko} the tomographic probability representation of spin (qudit) states was introduced.
In this representation the qudit states are identified with the spin-tomogram which is fair probability distribution function determined
by the density operator of the states. The relation of the density operator to the spin-tomogram is invertible. Due to this the tomogram contains
the complete information on the qudit state. For several qudits the spin-tomogram is also determined as the state density operator
and it is a joint probability distribution which provides the possibility to reconstruct the density operator.
Since the qudit state in the tomographic probability representation is identified with the standard probability distribution one
can use all the characteristics of the distributions like Shannon entropy and information as well as other entropies \cite{Renyi,Tsallis}.
The von Neumann entropy was shown \cite{MankoMendes} to be the minimum of the spin-tomographic Shannon entropy with respect to all the unitary transforms in
Hilbert space of the qudit system.
There exist different kinds of entropic inequalities for both classical  and quantum systems \cite{Lieb,Winter,Ruskai,Petz,Rastegin}.
The inequalities relating spin-tomographic and von Neumann entropies were used both for composite and noncomposite systems in \cite{Chernega,Chernega:14,Bregens,OlgaMankoarxiv}. The particular quantum state which has properties to be either separable or entangled depending on the parameter values of
 its density  matrix is the Werner state \cite{Werner} of two qubits.
\par  The aim of our work is to study the tomographic Shannon and von Neumann  entropies and informations discussed \cite{OlgaMankoarxiv}
on the example of the Werner state. We discuss the quantum correlations in the state using the specific characteristics of two-qubit density matrix.
This characteristics is the difference of quantum von Neumann information and maximum of the Shannon tomographic information taken with respect to all the
local unitary transforms in the Hilbert space of this bipartite qubit systems. We calculate explicitly the characteristics and analyze   this parameter behavior as
function of Werner state parameters.
\par The paper is organized as follows. In Sec.~\ref{sec.1} we review  the tomographic probability representation example of Werner state and introduce the tomographic
Shannon information and entropy for this two-qubit state. In Sec.~\ref{sec.2} we discuss the maximum of the spin-tomographic entropy of the composite two-qubit system with respect to the local unitary transforms in the Hilbert space.
\section{ Entropy and information for the  Werner state}\label{sec.1}
The tomographic probability distribution for spin states
provides the possibility to describe the states with density matrix $\rho$ of two qubits by means of tomogram.
By definition the spin tomogram is
\begin{eqnarray}\label{8}\omega(m_1,m_2,\overline{n}_1,\overline{n}_2)&=&\langle m_1,m_2|U\cdot\rho\cdot U^{\dagger}|m_1,m_2\rangle.
\end{eqnarray}
Here $m_{1,2}=-j,-j+1,\ldots,j$, $j= 0,1/2,1\ldots$ are spin projections and $U$ is the rotation matrix
\begin{eqnarray}\label{9}U&=&\left(
                       \begin{array}{cc}
                         \cos\frac{\theta_1}{2} e^{\frac{i(\varphi_1+\psi_1)}{2}}& \sin\frac{\theta_1}{2} e^{\frac{i(\varphi_1-\psi_1)}{2}} \\
                         -\sin\frac{\theta_1}{2}e^{\frac{i(\psi_1-\varphi_1)}{2}} & \cos\frac{\theta_1}{2} e^{\frac{-i(\varphi_1+\psi_1)}{2}}\\
                       \end{array}
                     \right)\otimes
                     \left(
                       \begin{array}{cc}
                         \cos\frac{\theta_2}{2} e^{\frac{i(\varphi_2+\psi_2)}{2}}& \sin\frac{\theta_2}{2} e^{\frac{i(\varphi_2-\psi_2)}{2}} \\
                         -\sin\frac{\theta_2}{2} e^{\frac{i(\psi_2-\varphi_2)}{2}} & \cos\frac{\theta_2}{2} e^{\frac{-i(\varphi_2+\psi_2)}{2}} \\
                       \end{array}
                     \right).
\end{eqnarray}
The matrix \eqref{9} is considered as the direct product of two matrices of irreducible representations of $SU(2)$ - group \cite{Man}.
The Werner state  of two qubits is determined by density matrix \cite{Werner} of the form
\begin{eqnarray}\label{10}\rho_{W}(p)&=&\left(
                                 \begin{array}{cccc}
                                   \rho_{1111}& \rho_{1112}& \rho_{1121}& \rho_{1122}\\
                                   \rho_{1211}& \rho_{1212}& \rho_{1221}& \rho_{1222}\\
                                   \rho_{2111}& \rho_{2112}& \rho_{2121}& \rho_{2122}\\
                                   \rho_{2211}& \rho_{2212}& \rho_{2221}& \rho_{2222}\\
                                 \end{array}
                               \right)
=\left(
                     \begin{array}{cccc}
                       \frac{1+p}{4} & 0 & 0 & \frac{p}{2}\\
                       0 & \frac{1-p}{4}& 0 & 0 \\
                       0 & 0 & \frac{1-p}{4} & 0 \\
                       \frac{p}{2} & 0 & 0 & \frac{1+p}{4} \\
                     \end{array}
                   \right),
\end{eqnarray}
where parameter $-\frac{1}{3}\leq p\leq1$. The parameter domain $\frac{1}{3}< p\leq1$ corresponds to the entangled state.
\par The eigenvalues of \eqref{10} are
\begin{eqnarray*}&&\lambda_1 = \frac{1+3p}{4}, \quad \lambda_{2,3,4} = \frac{1-p}{4}.
\end{eqnarray*}
The reduced density matrices of the first and the second qubit are the following
\begin{eqnarray*}\rho_1&=&\left(
                              \begin{array}{cc}
                                \rho_{1111}+\rho_{1212} & \rho_{1121}+\rho_{1222} \\
                               \rho_{2111}+\rho_{2212} &  \rho_{2121}+\rho_{2222} \\
                              \end{array}
                            \right)
=\left(
                            \begin{array}{cc}
                              \frac{1}{2} & 0 \\
                              0 & \frac{1}{2} \\
                            \end{array}
                          \right),\\
                          \rho_2&=&\left(
                              \begin{array}{cc}
                                \rho_{1111}+\rho_{2121} & \rho_{1112}+\rho_{2122} \\
                               \rho_{1211}+\rho_{2221} &  \rho_{1212}+\rho_{2222} \\
                              \end{array}
                            \right)=\left(
                            \begin{array}{cc}
                              \frac{1}{2} & 0 \\
                              0 & \frac{1}{2} \\
                            \end{array}
                          \right).
\end{eqnarray*}
Hence the von Neumann entropies of both qubit states and the entropy of the whole system are
\begin{eqnarray}\label{11}S_1 &=& -Tr\rho_1\ln\rho_1 = \ln2, \quad S_2 =-Tr\rho_2\ln\rho_2= \ln2,\\ \nonumber
S_{12} &=&-Tr\rho(p) ln\rho(p)=-\frac{1+3p}{4}\ln\left(\frac{1+3p}{4}\right)-3\frac{1-p}{4}\ln\left(\frac{1-p}{4}\right).
\end{eqnarray}
The quantum information is defined as the difference of the sum of the entropies of the first and the second qubit states and the entropy of the two-qubit state, i.e.
\begin{eqnarray}\label{12}I_q&=&S_1 +S_2-S_{12}.
\end{eqnarray}
Obviously, the quantum information satisfies the inequality $I_q\geq 0$. To construct the state tomogram we have to calculate the diagonal matrix elements of the density matrix
in unitary rotated basis in system Hilbert space.
\par The diagonal matrix elements of the matrix $U\cdot\rho\cdot U^{\dagger}$ are the following
\begin{eqnarray}\label{7}\omega_{11}(\uparrow,\uparrow)&=&\frac{1}{4}\left(p\left(\cos\theta_1\cos\theta_2 + \cos(\psi_1 + \psi_2)\sin\theta_1\sin\theta_2\right) + 1\right),\\\nonumber
\omega_{22}(\uparrow,\downarrow)&=&\frac{1}{4}\left(1 - p\left(\cos(\psi_1 + \psi_2)\sin\theta_1\sin\theta_2 + \cos\theta_1\cos\theta_2\right)\right),\\\nonumber
\omega_{33}(\downarrow,\uparrow)&=&\frac{1}{4}\left(1 - p\left(\cos(\psi_1 + \psi_2)\sin\theta_1\sin\theta_2 + \cos\theta_1\cos\theta_2\right)\right),\\\nonumber
\omega_{44}(\downarrow,\downarrow)&=&\frac{1}{4}\left(p\left(\cos\theta_1\cos\theta_2 + \cos(\psi_1 + \psi_2)\sin\theta_1\sin\theta_2\right) + 1\right).
\end{eqnarray}
Above we introduced the notations for the tomographic probabilities given by equation \eqref{8}, for example $\omega_{11}(\uparrow,\uparrow)\equiv\omega\left(+\frac{1}{2},+\frac{1}{2},\overline{n}_1,\overline{n}_2\right)$.
It is easy to verify that the  trace of the rotated density matrix satisfies the normalization condition
\begin{eqnarray*}&&Tr\left(U\cdot\rho\cdot U^{\dagger}\right)=\omega_{11}(\uparrow,\uparrow)+\omega_{22}(\uparrow,\downarrow)+
\omega_{33}(\downarrow,\uparrow)+\omega_{44}(\downarrow,\downarrow)=1.
\end{eqnarray*}
Marginal distributions corresponding to the first and the second qubit density matrix are
\begin{eqnarray*}W_1(\uparrow,\overline{n}_1)&=&\omega_{11}(\uparrow,\uparrow)+\omega_{22}(\uparrow,\downarrow),\quad
W_1(\downarrow,\overline{n}_1)=\omega_{33}(\downarrow,\uparrow)+\omega_{44}(\downarrow,\downarrow),\\
W_2(\uparrow,\overline{n}_2)&=&\omega_{11}(\uparrow,\uparrow)+\omega_{33}(\downarrow,\uparrow),\quad
W_2(\downarrow,\overline{n}_2)=\omega_{22}(\uparrow,\downarrow)+\omega_{44}(\downarrow,\downarrow).
\end{eqnarray*}
Thus, according to definition of Shannon entropy \cite{Shannon} we can construct the following tomographic entropies of the qubit subsystems
\begin{eqnarray}\label{1}H_1 &=& -W_1(\uparrow,\overline{n}_1)\ln W_1(\uparrow,\overline{n}_1)-W_1(\downarrow,\overline{n}_1)\ln W_1(\downarrow,\overline{n}_1)=\ln2,\\ \nonumber
H_2&=& -W_2(\uparrow,\overline{n}_2)\ln W_2(\uparrow,\overline{n}_2)-W_2(\downarrow,\overline{n}_2)\ln W_2(\downarrow,\overline{n}_2)=\ln2.
\end{eqnarray}
The tomographic Shannon entropy  of the bipartite system reads
\begin{eqnarray}\label{1_1}
H_{12}&=&-\omega_{11}(\uparrow,\uparrow)\ln\omega_{11}(\uparrow,\uparrow) - \omega_{22}(\uparrow,\downarrow)\ln\omega_{22}(\uparrow,\downarrow)-
\omega_{33}(\downarrow,\uparrow)\ln\omega_{33}(\downarrow,\uparrow) - \omega_{44}(\downarrow,\downarrow)\ln\omega_{44}(\downarrow,\downarrow).
\end{eqnarray}
We define the information $I_t$ as maximum of the sum of the difference between the sum of entropies \eqref{1} of subsystems and  the entropy of the whole system \eqref{1_1}
\begin{eqnarray}\label{5}I_t &=& \max\limits_{\psi_1,\psi_2,\theta_1,\theta_2}(H_1+H_2-H_{12})
\end{eqnarray}
and it satisfies the inequality $I_t\geq0$.
\section{Maximum of the Shannon information}\label{sec.2}
Let us introduce the following notation $\widetilde{H} \equiv \widetilde{H}(\psi_1,\psi_2,\theta_1,\theta_2,p)=H_1+H_2-H_{12}$. Using \eqref{1}, \eqref{1_1} and \eqref{7} it is straightforward to verify that
\begin{eqnarray}\label{6}
\widetilde{H}
&=&\ln4-\frac{1}{2}\ln\left(\frac{1}{4}\left(1 - p\cos(\psi_1 + \psi_2)\sin\theta_1\sin\theta_2 - p\cos\theta_1\cos\theta_2\right)\right)\\ \nonumber
&\cdot&\left(p\cos\theta_1\cos\theta_2 + p\cos(\psi_1 + \psi_2)\sin\theta_1\sin\theta_2 - 1\right) \\ \nonumber
&+& \frac{1}{2}\ln\left(\frac{1}{4}\left(p\cos\theta_1\cos\theta_2 + p\cos(\psi_1 + \psi_2)\sin\theta_1\sin\theta_2 + 1\right)\right)\\ \nonumber
&\cdot&\left(p\cos\theta_1\cos\theta_2 + p\cos(\psi_1 + \psi_2)\sin\theta_1\sin\theta_2 + 1\right).
\end{eqnarray}
To find the maximum of $\widetilde{H}$  with respect to angles $\psi_1,\psi_2,\theta_1,\theta_2$ we must first find its stationary points. Hence, taking the first derivatives
\begin{eqnarray*}\frac{\partial(\widetilde{H})}{\partial\theta_1} &=&\frac{p}{2}(\cos\theta_2\sin\theta_1 - \cos(\psi_1 + \psi_2)\cos\theta_1\sin\theta_2)\\
&\cdot&\Bigg(\ln\left(\frac{1}{4}\left(1-p\cos(\psi_1 + \psi_2)\sin\theta_1\sin\theta_2 - p\cos\theta_1\cos\theta_2\right)\right)\\
& -& \ln\left(\frac{1}{4}\left(p\cos\theta_1\cos\theta_2 + p\cos(\psi_1 + \psi_2)\sin\theta_1\sin\theta_2 + 1\right)\right)\Bigg),\\
\frac{\partial(\widetilde{H})}{\partial\theta_2} &=&\frac{p}{2}
(\cos\theta_1\sin\theta_2 - \cos(\psi_1 + \psi_2)\cos\theta_2\sin\theta_1)\\
&\cdot&\Bigg(\ln\left(\frac{1}{4}\left(1 - p\cos(\psi_1 + \psi_2)\sin\theta_1\sin\theta_2 - p\cos\theta_1\cos\theta_2\right)\right)\\
 &-& \ln\left(\frac{1}{4}\left(p\cos\theta_1\cos\theta_2 + p\cos(\psi_1 + \psi_2)\sin\theta_1\sin\theta_2+1\right)\right)\Bigg),\\
\frac{\partial(\widetilde{H})}{\partial\psi_1}&=&\frac{\partial(\widetilde{H})}{\partial\psi_2}
=\frac{p}{2}\sin(\psi_1 + \psi_2)\sin\theta_1\sin\theta_2\\
&\cdot&\Bigg(\ln\left(\frac{1}{4}\left(1 - p\cos(\psi_1 + \psi_2)\sin\theta_1\sin\theta_2-p\cos\theta_1\cos\theta_2\right)\right)\\
 &-& \ln\left(\frac{1}{4}\left(p\cos\theta_1\cos\theta_2 + p\cos(\psi_1 + \psi_2)\sin\theta_1\sin\theta_2+1\right)\right)\Bigg)
\end{eqnarray*}
and equating them to zero we can obtain that the critical points $\Theta^0=(\theta_1^0,\theta_2^0,\psi_1^0,\psi_2^0)$ are
\begin{itemize}
                     \item $\theta_1=\theta_2=\pi n, n=0,1,\ldots $ for $\forall \psi_1,\psi_2$,
                     \item $\theta_1=\pi/2+\pi n, \theta_2=\pi n, n=0,1,\ldots$ for $\forall \psi_1,\psi_2$,
                     \item $\theta_1=\pi n,\theta_2=\pi/2+\pi n, n=0,1,\ldots$ for $\forall \psi_1,\psi_2$,
                     \item $\theta_1=\theta_2=\pi/2+\pi n, n=0,1,\ldots$ for $\psi_1+\psi_2=\pi m$ or $\psi_1+\psi_2=\pi/2+\pi m$, $m=0,1,\ldots$,
                     \item $\theta_1=\pi/2+\pi n, \psi_1+\psi_2=\pi/2+\pi m$ for $\forall\theta_2$, $n,m=0,1,\ldots$,
                     \item $\theta_2=\pi/2+\pi n, \psi_1+\psi_2=\pi/2+\pi m$ for $\forall\theta_1$, $n,m=0,1,\ldots$
                   \end{itemize}
The second differential can be written in a quadratic form $d^2\widetilde{H}(\Theta)$ with  determinant
  \begin{eqnarray}\label{2}&&\begin{array}{|ccc|}
                        \frac{\partial^2(\widetilde{H})}{\partial\theta_1^2}& \frac{\partial^2(\widetilde{H})}{\partial\theta_2\partial\theta_1} & \frac{\partial^2(\widetilde{H})}{\partial\psi_1\partial\theta_1}  \\
                        \frac{\partial^2(\widetilde{H})}{\partial\theta_1\partial\theta_2}& \frac{\partial^2(\widetilde{H})}{\partial\theta_2^2} &  \frac{\partial^2(\widetilde{H})}{\partial\psi_1\partial\theta_2}    \\
                       2\frac{\partial^2(\widetilde{H})}{\partial\theta_1\partial\psi_1}&   2\frac{\partial^2(\widetilde{H})}{\partial\theta_2\partial\psi_1} & 2\frac{\partial^2(\widetilde{H})}{\partial\psi_1^2},
                     \end{array}
\end{eqnarray}
where we noticed that  $\frac{\partial^2(\widetilde{H})}{\partial\theta\partial\psi_1}=\frac{\partial^2(\widetilde{H})}{\partial\theta\partial\psi_2}$. According to a sufficient condition for an extremum if $d^2\widetilde{H}(\Theta^0)$ - is negatively defined quadratic form then $\Theta^0$ is a strict maximum of the function $\widetilde{H}(\psi_1,\psi_2,\theta_1,\theta_2,p)$. By Sylvester's criterion, if all of the leading principal minors of \eqref{2} are negative, then the quadratic form $d^2\widetilde{H}(\Theta^0)$ is negative.
\par For example we can take $\theta_1=\theta_2=\pi/2+\pi n, n=0,1,\ldots$. Then determinant \eqref{2} for $\psi_1+\psi_2=\pi m$, $m=0,1,\ldots$ is
\begin{eqnarray*}&&\begin{array}{|ccc|}
                     \frac{p}{2}\left(\ln\left(\frac{1-p}{4}\right) - \ln\left(\frac{1+p}{4}\right)\right) & \frac{p}{2}\left(\ln\left(\frac{1-p}{4}\right) - \ln\left(\frac{1+p}{4}\right)\right) & 0 \\
                     \frac{p}{2}\left(\ln\left(\frac{1-p}{4}\right) - \ln\left(\frac{1+p}{4}\right)\right) & \frac{p}{2}\left(\ln\left(\frac{1-p}{4}\right) - \ln\left(\frac{1+p}{4}\right)\right) & 0\\
                     0 & 0 & p\left(\ln\left(\frac{1-p}{4}\right) - \ln\left(\frac{1+p}{4}\right)\right),
                   \end{array}
\end{eqnarray*}
and for $\psi_1+\psi_2=\pi/2+\pi m$ it is
\begin{eqnarray*}&&\begin{array}{|ccc|}
                     0 & 0 & 0 \\
                     0 & 0 & 0 \\
                     0 & 0 & 2p^2
                   \end{array}
\end{eqnarray*}
It is straightforward to verify that both determinants are equal to zero. Thus, $\Theta^0_1=(\pi/2+\pi n,\pi/2+\pi n \psi_1+\psi_2=\pi m$ or $\psi_1+\psi_2=\pi/2+\pi m$, $n,m=0,1,\ldots$ is not an extremum point.
Similarly, for all other stationary points it can be proved that the second differential \eqref{2} becomes zero. Hence there is no global extremum of the function $\widetilde{H}(\psi_1,\psi_2,\theta_1,\theta_2,p)$.
\par Due to the form of stationary points it is clear that we can find $\theta_1^0,\theta_2^0$ that maximize $\widetilde{H}(\psi_1,\psi_2,\theta_1,\theta_2,p)$ with  fixed angles $\psi_1+\psi_2=\pi m$ or $\psi_1+\psi_2=\pi/2+\pi m$, $m=0,1,\ldots$.
\par
\begin{figure}[ht]
\begin{center}
\begin{minipage}[ht]{0.49\linewidth}
\includegraphics[width=1\linewidth]{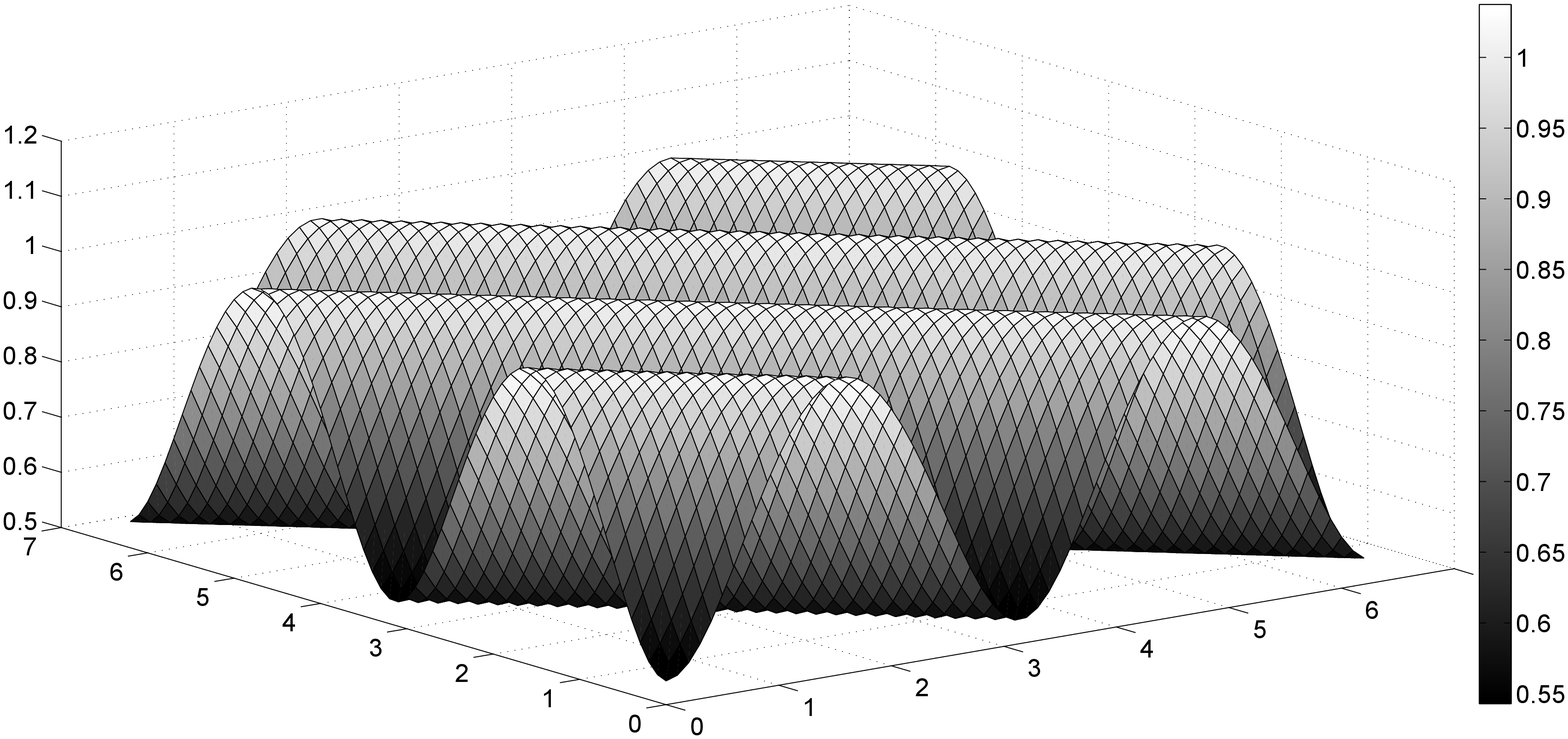}
\vspace{-4mm}
\caption{Difference \eqref{3} for fixed angles $\psi_1,\psi_2$ and $p=0.9$}
\label{fig:3}
\end{minipage}
\hfill
\begin{minipage}[ht]{0.49\linewidth}
\includegraphics[width=1\linewidth]{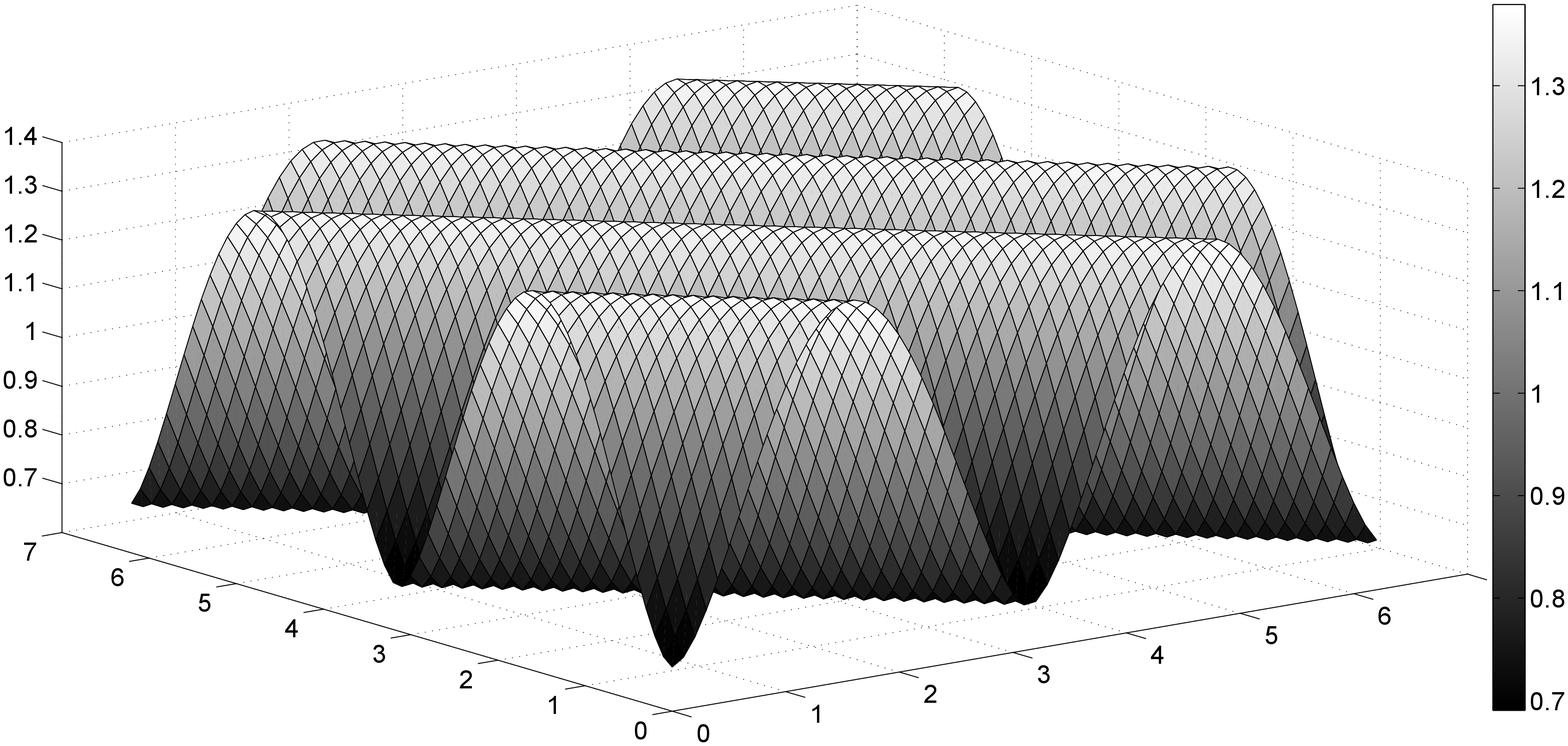}
\vspace{-4mm}
\caption{Difference \eqref{3} for fixed angles $\psi_1,\psi_2$ and $p=0.999$}
\label{fig:4}
\end{minipage}
\end{center}
\end{figure}

 The difference of quantum information $I_q$ and maximum
of the unitary tomographic information $I_t$ is
\begin{eqnarray}\label{3}&&I_q -I_t = \triangle I\geq0.
\end{eqnarray}
For fixed angles $\psi_1,\psi_2$ this difference is shown in Figure \ref{fig:3} for $p=0.9$ and in Figure \ref{fig:4} for $p=0.999$.
It is visible that with increasing of parameter $p$ the minimal value of difference \eqref{3} increases. In Figure \ref{fig:3} the minimal value of \eqref{3} is about $0.65$ and in Figure \ref{fig:4} is about $0.7$. Let us find its limit value as $p\rightarrow 1$.
\par
\begin{figure}[ht]
\begin{center}
\begin{minipage}[ht]{0.60\linewidth}
\includegraphics[width=1\linewidth]{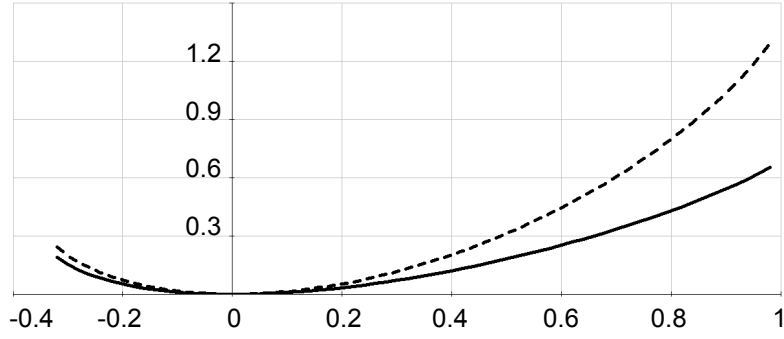}
\vspace{-4mm}
\caption{$I_q -I_t$ Solid line: $(\theta_1=\pi,\theta_2=\pi,\psi_1+\psi_2=\pi m)$, $(\theta_1=\pi/2,\theta_2=\pi/2,\psi_1+\psi_2=\pi m)$, $(\theta_1=\pi,\theta_2=\pi,\psi_1+\psi_2=\pi/2+\pi m)$. Dashed line: $(\theta_1=\pi/2,\theta_2=\pi,\psi_1+\psi_2=\pi m)$, $(\theta_1=\pi,\theta_2=\pi/2,\psi_1+\psi_2=\pi m)$,
$(\theta_1=\pi/2,\theta_2=\pi,\psi_1+\psi_2=\pi/2+\pi m)$, $(\theta_1=\pi,\theta_2=\pi/2,\psi_1+\psi_2=\pi/2+\pi m)$, $(\theta_1=\pi/2,\theta_2=\pi/2,\psi_1+\psi_2=\pi/2+\pi m)$ }
\label{fig:1}
\end{minipage}
\end{center}
\end{figure}

\begin{figure}[ht]
\begin{center}
\begin{minipage}[ht]{0.70\linewidth}
\includegraphics[width=1\linewidth]{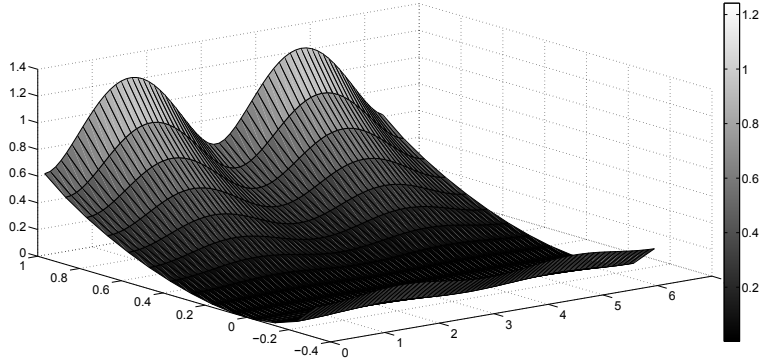}
\vspace{-4mm}
\caption{$I_q -I_t$ for $\psi_1+\psi_2=\pi$, $-1/3<p<1$ for $\theta_1\in[0,2\pi] $}
\label{fig:5}
\end{minipage}
\end{center}
\end{figure}

 To this end we use the well-known relation $\lim\limits_{x\rightarrow0}x\ln x=0$. Hence, we obtain that \eqref{11} and \eqref{12} are
\begin{eqnarray*}\lim\limits_{p\rightarrow1}S_{12}&=&0,\quad \lim\limits_{p\rightarrow -1/3}S_{12}=-\ln3\approx-1.098612,\\
\lim\limits_{p\rightarrow1}I_q&=&\ln4,\quad \lim\limits_{p\rightarrow-1/3}I_q=\ln4-\ln3\approx 0.287682.
\end{eqnarray*}
For $\psi_1+\psi_2=\pi/2+\pi m$ and $(\theta_1,\theta_2)=(\pi,\pi)$ the Shannon entropy \eqref{6} can be rewritten as
\begin{eqnarray*}&&\widetilde{H}(p)=\ln4 - \ln(1/4 - p/4)(p/2 - 1/2) + \ln(p/4 + 1/4)(p/2 + 1/2)
\end{eqnarray*}
Then its limits are
\begin{eqnarray*}&&\lim\limits_{p\rightarrow1}\widetilde{H}(p)=\ln4-\ln2,\quad \lim\limits_{p\rightarrow-1/3}\widetilde{H}(p)=\frac{5}{3}\ln2+\ln3.
\end{eqnarray*}
Hence the limit values of \eqref{3} are
\begin{eqnarray*}&&\lim\limits_{p\rightarrow1}(I_q -I_t)= \ln2\approx0.693147,\quad \lim\limits_{p\rightarrow-1/3}(I_q -I_t) =\frac{1}{3}\ln2\approx0.231049.
\end{eqnarray*}
For $\psi_1+\psi_2=\pi/2+\pi m$ and $(\theta_1,\theta_2)=(\pi,\pi/2)$  entropy \eqref{6} is
\begin{eqnarray*}&&\lim\limits_{p\rightarrow1}\widetilde{H}(p)=0,\quad \lim\limits_{p\rightarrow-1/3}\widetilde{H}(p)=0
\end{eqnarray*}
and the limit values of \eqref{3} are
\begin{eqnarray*}&&\lim\limits_{p\rightarrow1}(I_q -I_t)= \ln4\approx1.386294,\quad \lim\limits_{p\rightarrow-1/3}(I_q -I_t) =\ln4-\ln3\approx 0.287682.
\end{eqnarray*}
These limits can be seen in  Figure \ref{fig:1} in all the stationary points with a varying $p$ and  additionally for the varying angle $\theta_1\in[0,2\pi] $ in
Figure \ref{fig:5}.
Hence, the minimum value of \eqref{3} is $\triangle I=\ln2$ as $p\rightarrow1$ and $\triangle I=\frac{1}{3}\ln2$ as $p\rightarrow-\frac{1}{3}$.
\section{Summary}
\pst
To conclude we point out the main results of the work.
We studied the correlations in Werner state
of two qubits.
The difference of von Neumann information $I_q$ and the maximum of tomographic information $I_t$
associated with correlations in the system must be nonnegative. This is shown in Figure \ref{fig:5} for a fixed  $\overline{n}_2=(\theta_2,\psi_2)$ and varying  latitude of $\overline{n}_1$. For $p\to1$ (maximally entanglement state) the difference tends to $\ln2$. The studied difference characterizes the degree of quantum correlations in the two-qubit system.
\section*{Acknowledgments}
\pst
L. A. M. acknowledges the financial support provided within the Russian Foundation for Basic Research, grant 13-08-00744 A.


\begin{thebibliography}{99}
\bibitem{schredinger:35}E. Schr\"{o}dinger, {\it Die gegenwartige Situation in der Quantenmechanik (The present situation in quantum mechanics)}, {\sl Naturwissenschaften}, \textbf{23}, 807--812; 823--828; 844--849 (1935).
\bibitem{Horn}J.F. Clauser, M.A. Horne, A. Shimony, R.A. Holt, {\sl Proposed experiment to test local hidden-variable theories}, {\sl Phys. Rev. Lett.}, \textbf{23 (15)}, 880--4 (1969).
\bibitem{Yurkevich} V.I. Man'ko, A. Yurkevich, {\it Tomographic Discord and Quantum Correlations in a System of Qubits}, {\sl J. Russ. Laser Res.}, \textbf{34 (5)}, 463--467 (2013).
\bibitem{Mscord} K. Modi, Aharon Brodutch, H. Cable, T. Paterek, and V. Vedral, {\it Quantum discord and other measures of quantum correlation}, {\sl  Rev. Mod. Phys.}, \textbf{84}, 1655-1707 (2012).
\bibitem{Shannon}C. E. Shannon,  {\it A Mathematical Theory of Communication}, {\sl Bell System Technical Journal}, \textbf{27}, 379 (1948).
\bibitem{Dodonov} V.V. Dodonov, V.I. Man'ko, {\it Positive distribution description for spin states}, {\sl Phys. Lett. A}, \textbf{229}, 335--339 (1997).
\bibitem{OVManko}V.I. Man'ko,  O.V. Man'ko, {\it Spin state tomography}, {\sl JETP}, \textbf{85 (3)}, 430 (1997).
\bibitem{Renyi}A. R\'{e}nyi, {\it On measures of information and entropy}, Proceedings of the fourth Berkeley Symposium on Mathematics, {\sl Statistics and Probability}, 547-561 (1961).
\bibitem{Tsallis}  C. Tsallis, {\it Possible generalization of Boltzmann-Gibbs statistics}, {\sl Journal of Statistical Physics}, \textbf{52}, 479--487 (1988).

\bibitem{Man} V.I. Man'ko and  L.A. Markovich, {\it Entropic inequalities and properties of some special functions},  {\sl J. Russ. Laser Res.}, \textbf{35 (2)}, (2014).

\bibitem{MankoMendes}R. Vilela Mendes, V.I. Man'ko, {\it On the problem of quantum control in infinite dimensions}, {\sl Journal of Physics A: Math. Theor.}, \textbf{44 (13)},
135302 (2011).
\bibitem{Lieb} E. H. Lieb, M. B. Ruskai, {\it Proof of the Strong Subadditivity of Quantum
Mechanical Entropy}, {\sl J. Math. Phys.}, \textbf{14}, 1938--1941, (1973).
\bibitem{Winter}S. Wehner, A. Winter, {\it Entropic uncertainty relations—a survey}, {\sl New J. Phys.}, \textbf{12}, 025009 (2010).
\bibitem{Ruskai}E.H. Lieb, M.B. Ruskai, {\it Some Operator Inequalities of the Schwarz Type}, {\sl Adv. Math},  \textbf{12}, 269--273 (1974).
\bibitem{Petz}M. Ohya, D. Petz, {\it Quantum entropy and its use}, {\sl Springer-Verlag}, Berlin, (1993).
\bibitem{Rastegin}Alexey E. Rastegin, {\it Fano type quantum inequalities in terms of q-entropies},  \textbf{11 (6)}, 1895--1910, (2012).
\bibitem{Chernega}V.N. Chernega, O.V. Man'ko, V.I. Man'ko, {\it Generalized quit portrait of the qutritstate density matrix}, {\sl J. Russ. Laser Res.}, \textbf{34 (4)}, 383--387 (2013).
\bibitem{Chernega:14}V.N. Chernega, O.V. Man'ko, {\it Tomographic and improved subadditivity conditions for two qubits and qudit with j = 3/2}, {\sl J. Russ. Laser Res.}, \textbf{35 (1)}, 27--38 (2014).
\bibitem{Bregens}M.A. Man'ko, V.I. Man'ko, {\it Quantum correlations expressed as information and entropic inequalities for composite and noncomposite systems}, {\sl J. Phys. Conf. Ser. in the Proceedings of the XVI Symposium "Symmetries in Science"}, (Bregenz, Austria, July 21--26, 2013)
\bibitem{OlgaMankoarxiv}V.N. Chernega, O.V. Man'ko, V.I. Man'ko, {\it Subadditivity condition for spin-tomograms and density matrices of arbitrary composite and noncomposite qudit systems}, http://arxiv.org/abs/1403.2233 (2014).
\bibitem{Werner}R.F. Werner, {\it Quantum states with Einstein-Podolsky-Rosen correlations admitting a hidden-variable model}, {\sl Phys. Rev. A}, \textbf{40}, 4277 (1989).
\bibitem{Manko}M.A. Man'ko and  V.I. Man'ko, arXiv:1312.6988 (2013), to appear in Physica Scripta (2014).


\end{thebibliography}
\end{document}